\documentclass[epj]{svjour}
\usepackage{graphicx}
\usepackage{here}
\begin{document}
\title{Dynamics of the near threshold $\eta$ meson production in proton-proton interaction}
\author{R.~Czy{\.z}ykiewicz\inst{1,2}, P.~Moskal\inst{1,2}, H.-H.~Adam\inst{3}, A.~Budzanowski\inst{4},
        E.~Czerwi\'nski\inst{1}, D.~Gil\inst{1}, D.~Grzonka\inst{2}, M.~Hodana\inst{1,2},
        M.~Janusz\inst{1,2}, L.~Jarczyk\inst{1}, B.~Kamys\inst{1}, A.~Khoukaz\inst{3},
        K.~Kilian\inst{2}, P.~Klaja\inst{1,2}, B.~Lorentz\inst{2}, W.~Oelert\inst{2},
        C.~Piskor-Ignatowicz\inst{1}, J.~Przerwa\inst{1,2}, B.~Rejdych\inst{1},
        J.~Ritman\inst{2}, T.~Sefzick\inst{2}, M.~Siemaszko\inst{5}, M.~Silarski\inst{1}, J.~Smyrski\inst{1},
        A.~T\"aschner\inst{3}, K.~Ulbrich\inst{6}, P.~Winter\inst{7}, M.~Wolke\inst{2},
        P.~W\"ustner\inst{2}, M.~J.~Zieli\'nski\inst{1}, W.~Zipper\inst{5}
}                     
\institute{Institute of Physics, Jagellonian University, 30-059 Cracow, Poland \and
           IKP \& ZEL, Forschungszentrum J\"ulich, 52425 J\"ulich, Germany \and
           IKP, Westf\"alische Wilhelms-Universit\"at, 48149 M\"unster, Germany \and
           Institute of Nuclear Physics, 31-342 Cracow, Poland \and
           Institute of Physics, University of Silesia, Katowice, Poland \and
           ISK, Rheinische Friedrich-Wilhelms-Universit\"at, 53115 Bonn, Germany \and
           Department of Physics, University of Illinois at Urbana-Champaign, Urbana, IL 61801, USA}
\date{Received: date / Revised version: date}
%
\abstract{
We present the results of measurements of the analysing power for
the $\vec{p}p\to pp\eta$ reaction at the excess energies of Q~=~10 and 36~MeV,
and interpret these results within the framework of the meson exchange models.
The determined values of the analysing power at both excess
energies are consistent with zero implying that the
$\eta$ meson is produced predominantly in s-wave. 
} 
\authorrunning{R.~Czy{\.z}ykiewicz, P.~Moskal et al.}
\titlerunning{Dynamics of the near threshold $\eta$ meson production in proton-proton interaction}
\maketitle
\section{Introduction}
\label{intro}
From the precise measurements of the total cross 
section for the $\eta$ meson production in the 
$pp\to pp\eta$ reaction in the close-to-threshold 
region~\cite{total1,total2,total3,total4,total5,total6,total7,total8} 
it was concluded~\cite{theor1,theor2,theor3,theor4,theor5,theor6,theor7,wilkin,nakayama}
that this process proceeds through the excitation 
of one of the protons to the S$_{11}$(1535)
state, which subsequently deexcites via the emission 
of the $\eta$ meson and a proton. 
However, there 
are plenty of possible scenarios of the 
excitation of S$_{11}$(1535) resonance. In fact, exchange of any 
of the $\pi, \eta, \omega$, or $\rho$ mesons may 
contribute to the resonance creation. Considering the 
cross sections itself doesn't answer the 
question which out of these mesons give the significant 
contribution to the production amplitude. 

Some constraints may be deduced from the investigations
of the isospin dependence of the total cross section
for the $NN\to NN\eta$ reaction~\cite{calen}.
The ratio of the total cross sections for the $\eta$ meson 
production in proton-neutron collisions to the analogous
cross section with proton-proton colliding in the initial 
state was found to be about 6.5 in the close-to-threshold 
region, which revealed strong isospin dependence of the 
production process. This means that the production 
of the $\eta$ meson with the total isospin I=0 in the 
initial channel exceeds the production with the 
isospin I=1 by a factor of 12, suggesting~\cite{wilkinratio}
that the isovector meson exchange - the $\pi$ or $\rho$ meson 
exchange - is the dominant process leading to the 
excitation of the S$_{11}$ resonance. However, the 
relative contributions of the pseudoscalar $\pi$ meson and vector
$\rho$ meson still remain to be determined. 

Here, the measurements of the polarization observables 
can assist, because the predictions of the one boson 
exchange models~\cite{wilkin,nakayama} with respect 
to the analysing power are sensitive to the type of 
the exchanged meson. 

COSY-11 collaboration performed 
two measurements of the analysing power function at 
the beam momenta of p$_{beam}=2.010$~GeV/c and 
2.085~GeV/c, which for the $\vec{p}p\to pp\eta$ reaction
correspond to the excess energies of Q=10 and 36~MeV, respectively. 
Here we would like to briefly summarize the results of these measurements
and present the main conclusions we could have drawn from our studies.

\section{Results}
\label{sec:1}

In the measurements the COSY-11 detection setup~\cite{brauksiepe,jurek,klaja} has been used, 
along with the vertically polarized proton beam, which polarization was 
flipped from cycle to cycle in order to reduce the systematic uncertainties. 
For the detailed description of the experimental aparatus, method of 
measurement and analysis, the reader is referred to~\cite{doktorat}.

The tested predictions of the analysing power of reference~\cite{wilkin}
were based on the assumption of the $\rho$ meson exchange dominance
and the proton asymmetries taken from the photoproduction of the $\eta$
meson~\cite{photo}. In the case of the calculations of reference~\cite{nakayama}
the exchanges of all mesons have been taken into account in the framework 
of the relativistic meson exchange model of hadronic interactions and it 
was found in this model that the contribution from the pion exchange is the 
dominant one. 

\begin{figure}[h]
\begin{center}
  \includegraphics[width=4.2cm]{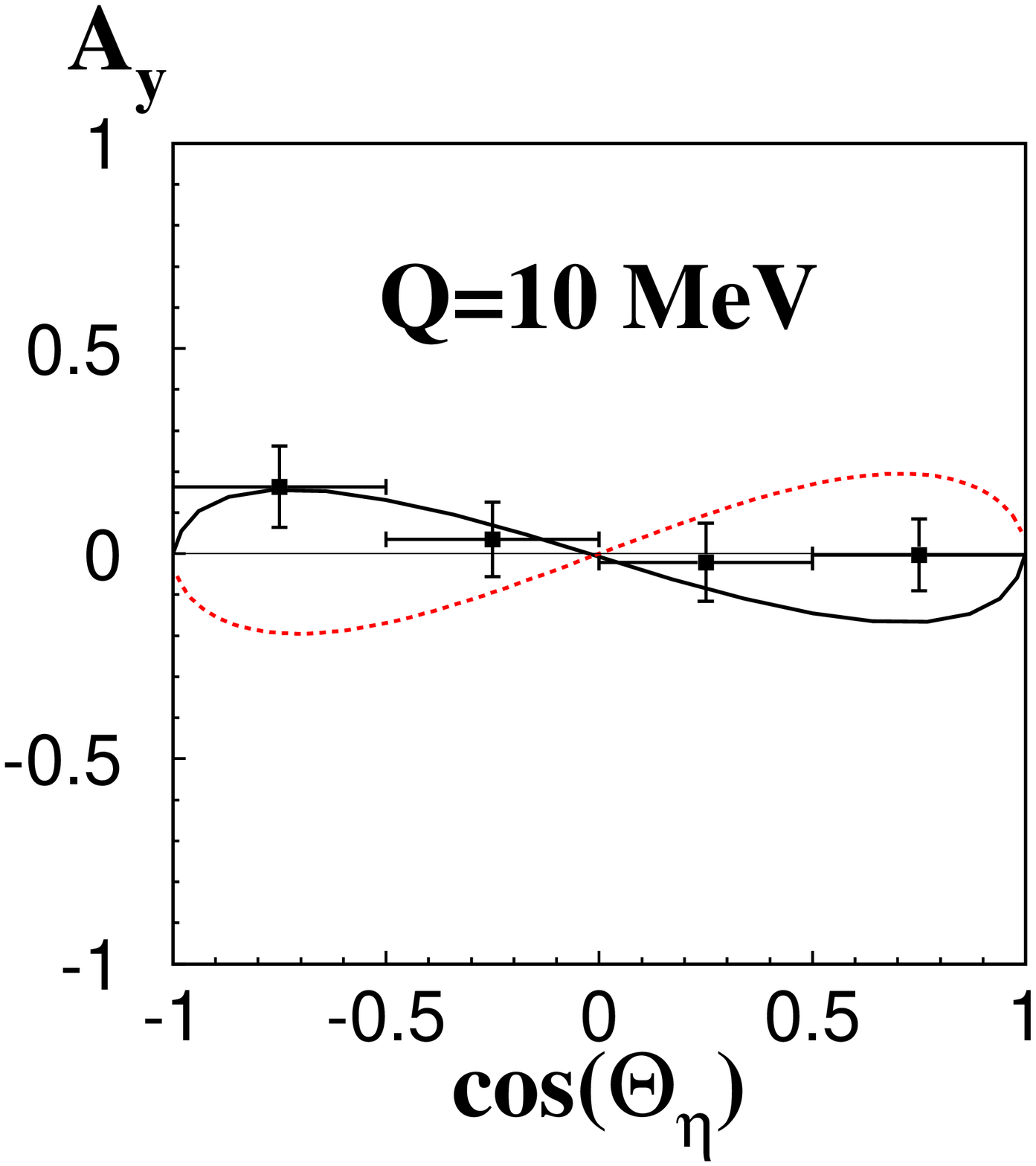}
  \includegraphics[width=4.2cm]{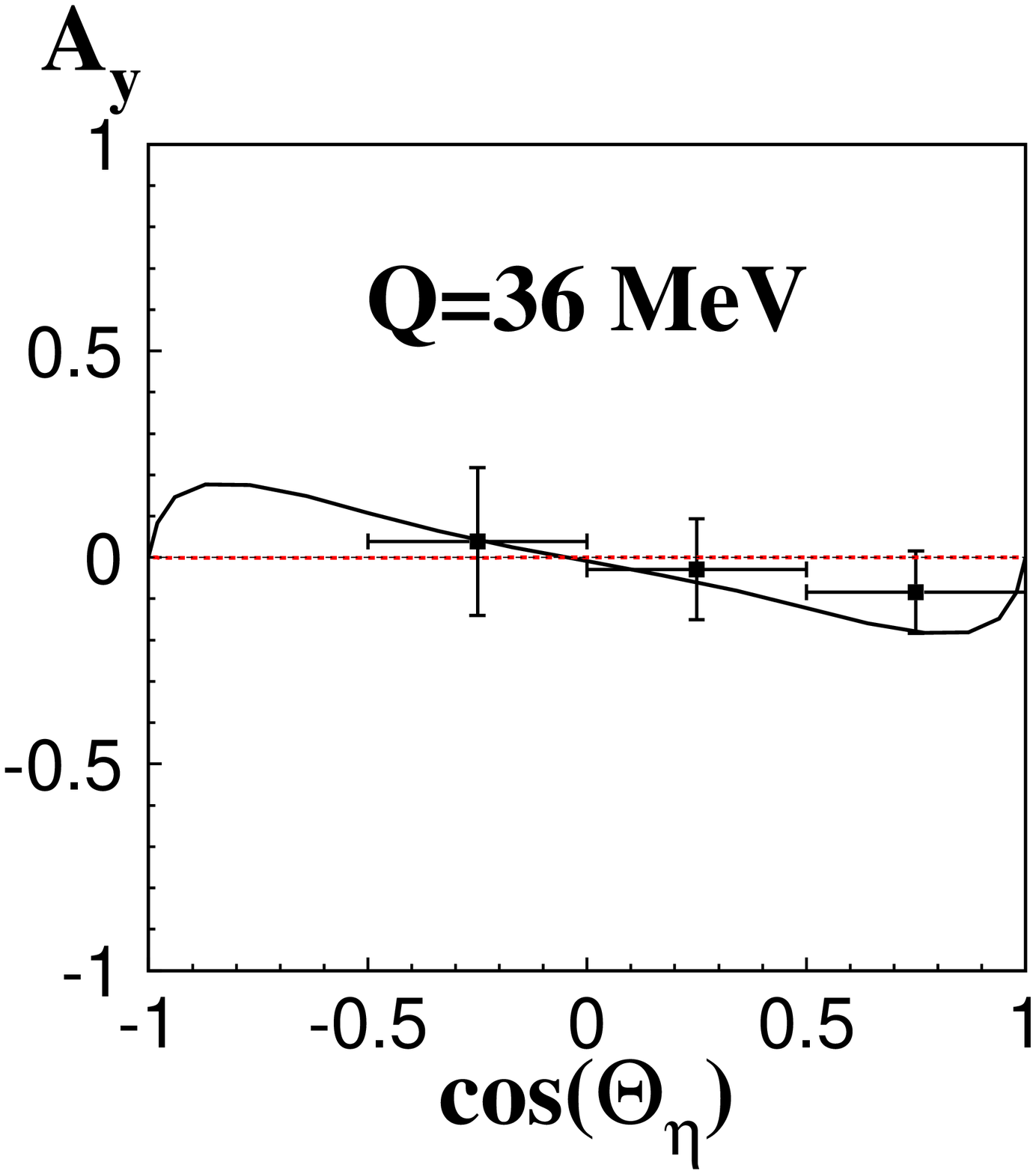}
\end{center}
\caption{ Analysing power A$_y$ for the $\vec{p}p\to pp\eta$ reaction
as a function of cosine of the polar angle of the $\eta$ meson 
production in the overall center-of-mass system for 
Q=10 MeV (left panel) and Q=36 MeV (right panel). Full 
lines are the predictions based on the pseudoscalar meson 
exchange model~\cite{nakayama} whereas the dotted lines represent
the calculations based on the vector meson exchange~\cite{wilkin}. 
In the right panel the dotted line is consistent with zero. 
Shown are the statistical uncertainties. 
}
\label{fig1}
\end{figure}

Fig.~\ref{fig1} shows the experimentally determined values of the 
analysing power A$_y$ for the $\vec{p}p\to pp\eta$ reaction
confronted with the theoretical predictions of the pseudoscalar~\cite{nakayama} 
and vector~\cite{wilkin} meson exchange dominance models. The $\chi^2$ test
of the correctness of these models have been performed and the 
reduced values of the $\chi^2$ were found to be equal to 0.54 (corresponding to the 
significance level of $\alpha_{psc}=0.81$) 
and 2.76 ($\alpha_{vec}=0.006$), for pseudoscalar and vector meson exchange dominance models, respectively.

\begin{figure}[h]
\begin{center}
  \includegraphics[width=4.2cm]{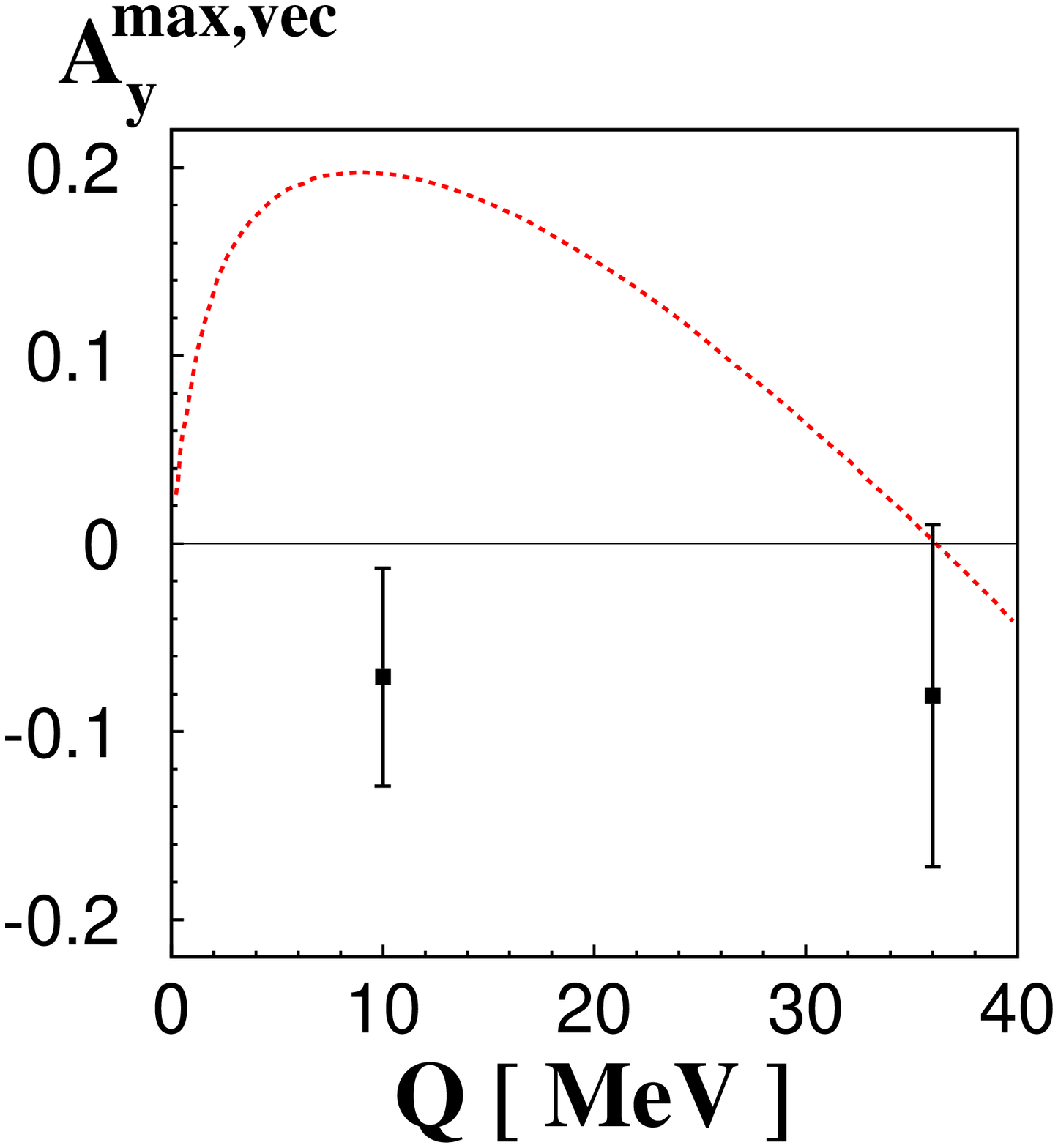}
  \includegraphics[width=4.2cm]{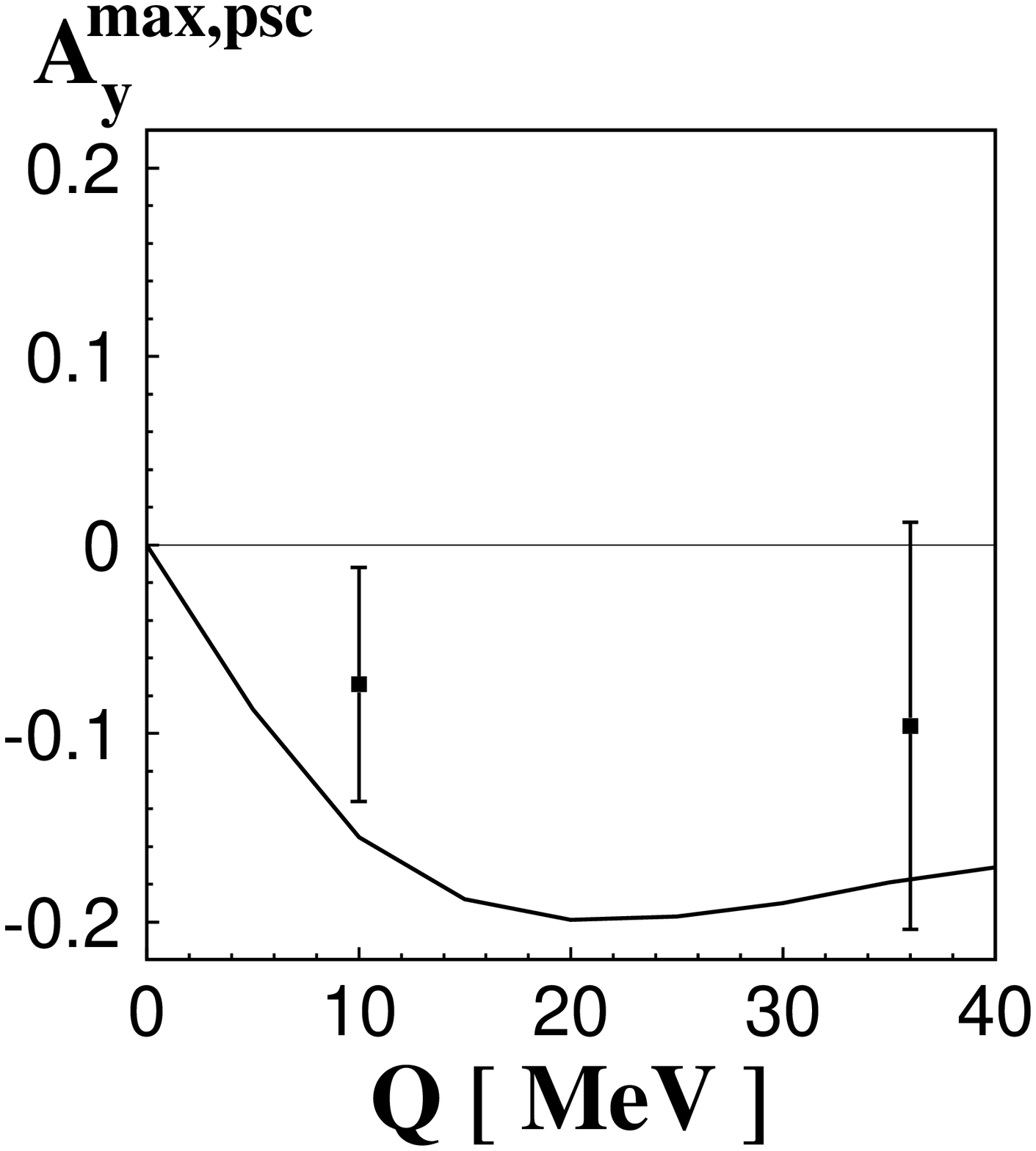}
\end{center}
\caption{Theoretical predictions for the energy dependence 
of the amplitude of A$_y^{max}$ confronted with the amplitudes 
determined experimentally at the excess energies of Q=10 and 37 MeV
for the vector (left) and pseudoscalar (right) meson exchange
dominance model.}
\label{fig2} 
\end{figure}

In the vector meson exchange dominance model~\cite{wilkin}
the angular distribution of the analysing power
is parameterized as a function of the polar angle of $\eta$ meson 
production in the center-of-mass system with the following equation:
\begin{equation}
A_y(\theta_{\eta}) = A_y^{max,vec} \sin{2\theta_{\eta}},
\label{aymax}
\end{equation}
where the amplitude A$_y^{max,vec}$ is a function of the excess energy Q,
shown as a dotted line in the left panel of Fig.~\ref{fig2}.

We have estimated the values of A$_y^{max,vec}$ comparing the experimental data
with predicted shape utilizing a $\chi^2$ test~\cite{doktorat,czyzyk1,czyzyk2}.
Determined experimental values are shown in Fig.~\ref{fig2} (left) along with the theoretical 
predictions according to the vector meson exchange dominance model~\cite{wilkin}. 
Analogously, the confrontation of the experimentally determined amplitude A$_y^{max,psc}$ with the 
predictions of the pseudoscalar meson exchange dominance model~\cite{nakayama}
are shown in Fig.~\ref{fig2} (right). Predictions of the model based on the $\pi$
mesons dominance are fairly consistent
with the data, whereas
the calculations based on the dominance of the $\rho$ meson exchange
differ from the data by more than four standard
deviations. However, the latter calculation used the proton
asymmetry ($T$) in eta photoproduction~\cite{photo}, within the
framework of the vector meson dominance model~\cite{sakurai}, as the basis of
their estimate. It should be noted that it has proved hard to
reconcile the experimental value of $T$ with the results of
photoproduction amplitude analyses~\cite{MAID}.

\section{Conclusions and outlook}

Taking into account the $\chi^2$ analysis of the
analysing power for the pseudoscalar and vector meson
exchange models we have shown that the predictions of the
pseudoscalar meson exchange dominance~\cite{nakayama} are in line with the
experimental data at the significance level of 0.81.
On the other hand, the assumption that the $\eta$ meson is produced solely via the
exchange of the $\rho$ meson~\cite{wilkin},
leads to the discrepancy between the theoretical predictions
and experimental data larger than four standard deviations.
It must be stated, however, that the production amplitude for
the $\rho$ meson exchange was determined based on the
vector meson dominance hypothesis and the photoproduction data~\cite{photo}.
At this point it is also worth mentioning that the recent calculations of the
$\eta$ meson production in the NN collisions performed in the
framework of the effective Lagrangian model~\cite{shyam} also indicate
the dominance of the pion exchange.

The analysing power values for both excess energies are
consistent with zero within one standard deviation. This is
in line with the results obtained by the DISTO~\cite{disto}
collaboration in the far-from-threshold energy region.
Such a result may indicate that the $\eta$ meson is
predominantly produced in the $s$-wave.

\section{Future perspectives}

Recently, the proposal for the measurement of the analyzing power function~\cite{hodana}
with the WASA-at-COSY aparatus~\cite{wasa} has been presented and awaits recommendation
of the COSY Programme Advisory Committee.
Measurements are planned with about 50 times better statistics which should enable
the error bars from Fig.~\ref{fig1} to be reduced of circa 7 times.

\begin{figure}[h]
\begin{center}
\includegraphics[width=4.2 cm]{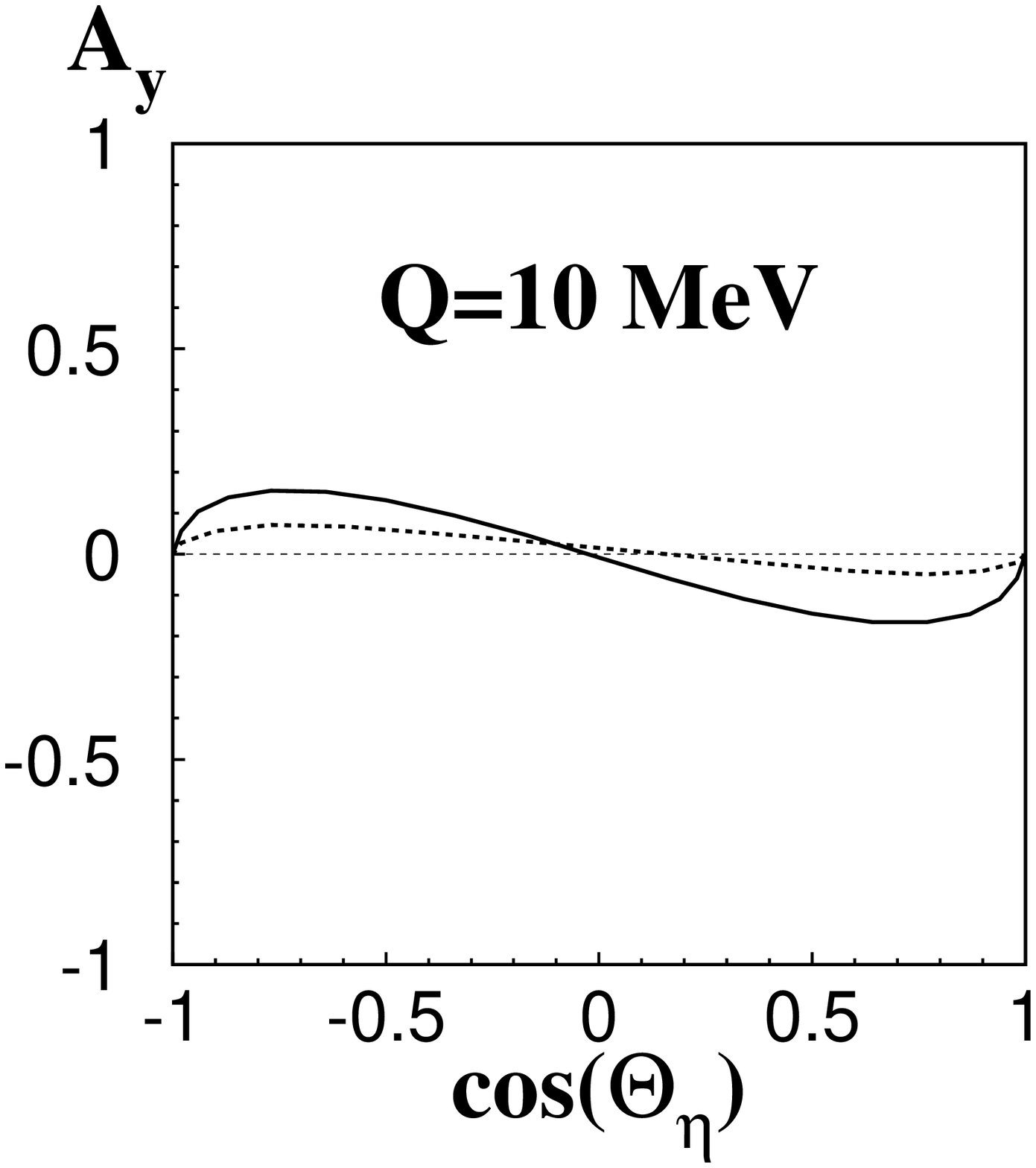}
\includegraphics[width=4.2 cm]{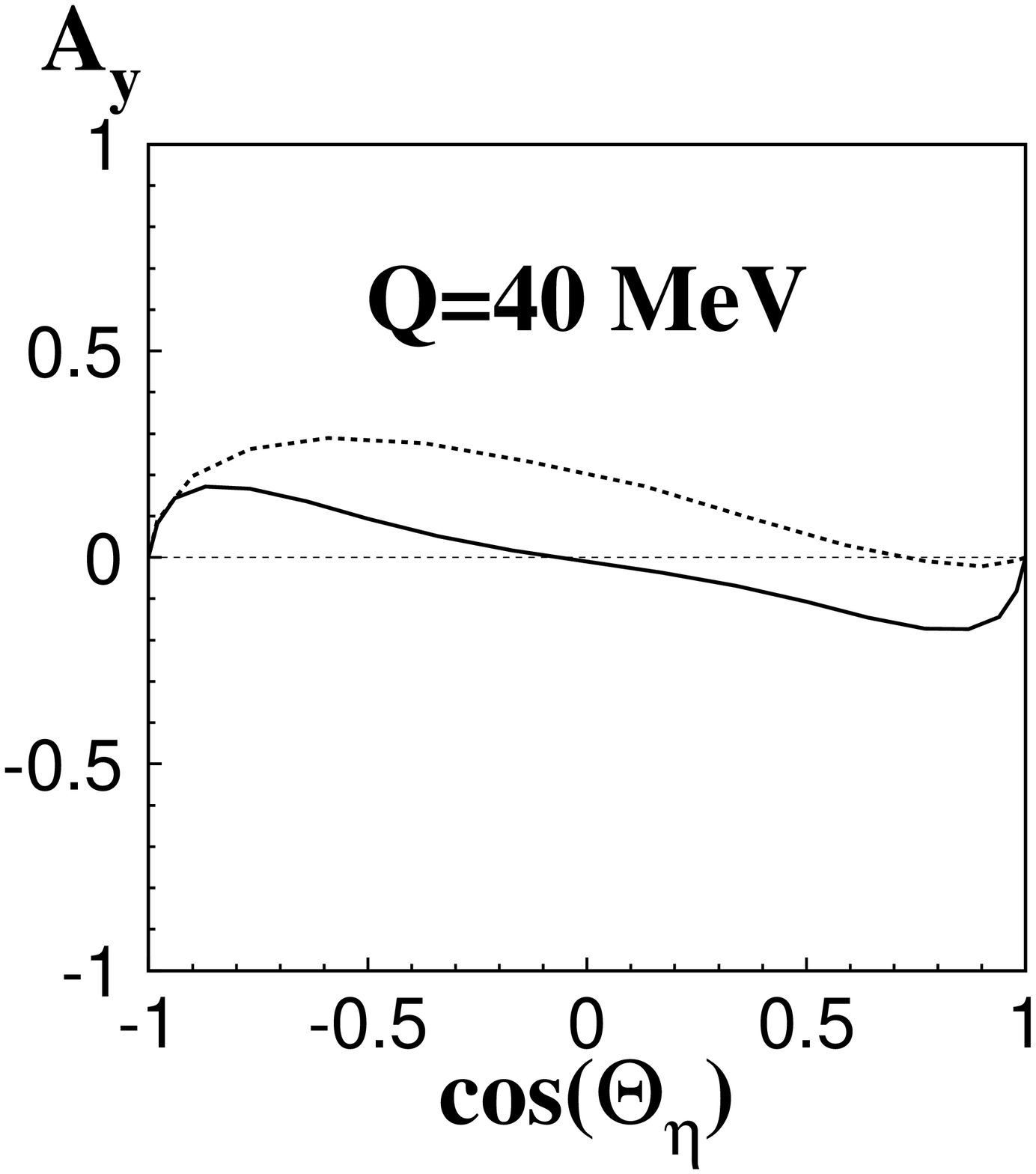}
\caption{ Predictions of the dependence of the analyzing power function
on the intermediate resonance type~\cite{kanzo-private}.
} \label{fig3}
\end{center}
\end{figure}

Fig.~\ref{fig3} presents the dependence of the analyzing power as a
function of the cosine of the polar angle of the $\eta$ meson emission
in the center-of-mass system on the intermediate resonance type~\cite{kanzo-private}.
Solid line are the calculations of the pseudoscalar meson exchange
model performed under assumption that only S$_{11}$(1535) resonance
contributes to the $\eta$ meson production amplitude, whereas the dotted line
represent the predictions of the same model, including D$_{13}$(1520), S$_{11}$(1535),
S$_{11}$(1650), and D$_{13}$(1700) resonances.
Therefore, the improvement in the measurement accuracy would enable to investigate
the influence of other-than-S$_{11}$(1535) resonances upon the production
amplitude.

Measurements of the analysing power A$_y$ with much higher statistics 
may also allow the model independent partial wave decomposition with an accuracy by far
better than resulting from the measurements of the distributions of the spin 
averaged cross sections. This is because the polarization observables 
can probe the interference terms between various partial amplitudes, even if they 
are negligible for the spin averaged distributions. More importantly, in case
of the $pp\to ppX$ reaction the interference terms between the transition with odd 
and even values of the angular momentum of the final state baryons are bound to vanish
for the cross sections~\cite{deloff,wilkinpriv}. This characteristic is due to the invariance
of all observables under the exchange of identical nucleons in the final state. Due to the same 
reason there is no interference between $s$ and $p$-waves of the $\eta$ meson in 
the differential cross sections~\cite{wilkinpriv}. However, $s$-$p$ interference does not 
vanish for the proton analysing power, and thus the precise measurements of A$_y$ could 
provide the first determination of the comparatively small $p$-wave contribution~\cite{wilkinpriv}, 
unreachable from spin averaged observables.

\vspace{1cm}

\section{Acknowledgment}
We acknowledge the support of the
European Community-Research Infrastructure Activity
under the FP6 programme (Hadron Physics, N4:EtaMesonNet,
RII3-CT-2004-506078), the support
of the Polish Ministry of Science and Higher Education under the grants
No. PB1060/P03/2004/26, 3240/H03/2006/31  and 1202/DFG/2007/03,
and  the support of the German Research Foundation (DFG) under the grant No. GZ: 436 POL 113/117/0-1.

\end{document}